\documentclass[structabstract]{aa}  
\usepackage{graphicx}
\begin{document}
   \title{The 2010 early outburst spectrum of the recurrent nova U Scorpii}


   \author{Styliani Kafka
          \inst{1}
          \and
          Robert Williams\inst{2}}

   \institute{Department of Terrestrial Magnetism, Carnegie Institution of Washington, 5241 Broad Branch Road NW, Washington, DC 20015, US\\
              \email{skafka@dtm.ciw.edu}
         \and
             Space Telescope Science Institute, 3700 San Martin Drive Baltimore, MD 21218.\\
             \email{wms@stsci.edu}
            }

   \date{Received; Accepted }

 
  \abstract
  {} 
   {We present optical spectra of the fast recurrent nova U Sco during its recent outburst, obtained within 24 hr of maximum light.}
   {We use medium resolution (R$\sim$4000) spectra taken with the with
     the MagE spectrograph on the Magellan (Clay) 6.5m telescope of
     the Las Campanas Observatories.}
   {The spectrum is notable for its lack of a low ionization transient heavy element absorption system that is visible in the large majority of novae near maximum light.  We suggest that this may be due to the dominance of inner Lagrangian L1 mass transfer and the absence of a circumbinary gas reservoir in this object}
   {}


   \maketitle
%

\section{Introduction}

Recurrent novae (RNe) are semi-detached binary systems consisting of  a massive white dwarf (WD; M$_{WD}$$\ge$1.2M$_{\sun}$) accreting material from a companion, with mass transfer rates of $\ge$10$^{-7}$M$_{\sun}$/year (\cite{2000ApJ...534L.189H}). With total system mass exceeding the Chandrasekhar mass limit, they are prime candidates for SNeIa progenitors, therefore they are monitored diligently. Their ``recurrent'' designation reflects the fact that they exhibit multiple nova eruptions on time
scales of several decades -- 
more frequent than classical novae whose eruptions have been observed only
once. Therefore they are excellent natural test beds for the study and
understanding of stellar explosions in accreting systems and, of course, the mechanism of mass accumulation on a near-Chandrasekhar limit WD. An excellent recent review on the properties of RNe can be found in \cite{2010ApJS..187..275S}.  

The object of this work, U Sco, is one of the most 
well-studied recurrent novae, not only for its short intra-outburst 
rate\footnote{The system has 10 recorded nova outbursts,
  starting in 1863, with an average inter-erruption interval of a
  decade; it also holds a record short inter-eruption interval of 7.88 years (1979 to 1987).}
but also because it is an eclipsing binary with an orbital period of 1.23 days
(Schaefer 1990, \cite{1995ApJ...447L..45S}). It consists of a massive
WD (M$_{WD}$$\ge$1.2M$_{\sun}$; \cite{2000AJ....119.1359A}) and a subgiant
mass-losing companion whose spectral type ranges between F8
(\cite{1992ApJ...396..267J}) and K2
(\cite{2000AJ....119.1359A}; also, Hanes 1985,
Schaefer 1990).
Because of its total mass exceeding the Chandrasekhar mass limit of 1.4 M$_{\sun}$, it is considered a favorable candidate for a SN Ia progenitor.
In quiescence, its optical magnitude ranges from V$\sim$18 (outside
eclipse) to V$\sim$19 in eclipse (Schaefer et al. 2010); the
maximum brightness in outburst reaches V$\sim$8.0 mag. Its rapid decline by 3 magnitudes from maximum light occurs in 4 days (figure 1) making it a very fast
recurrent nova. The principal mechanism leading to the eruption is
believed to be a thermonuclear runaway on the mass-accumulating WD, enriching the ISM with CNO-enhanced material.

This short communication presents one of the earliest  U Sco 2010 outburst spectra, obtained at moderate spectral resolution within  24 hours of discovery and peak brightness and showing  strong, wide Balmer lines with fast variability and blueshifted/redshifted velocities reaching 3000 km/sec.


\section{Observations}

\begin{figure}
   \centering
\includegraphics[angle=0,width=7cm]{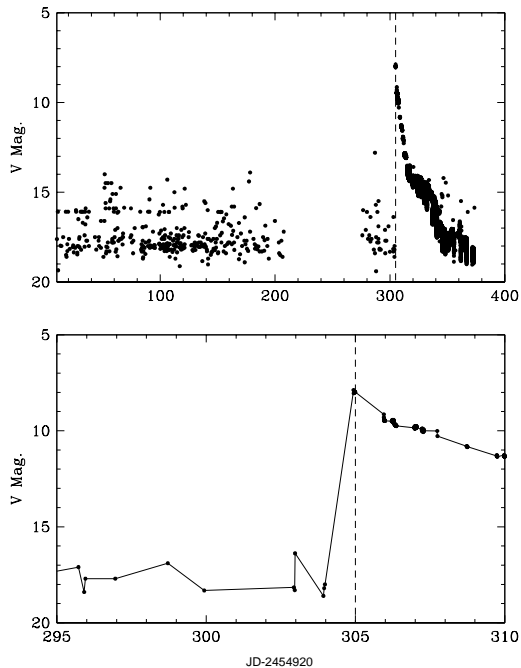}
   \caption{AAVSO light curve of U Sco, with the time of our
  spectroscopic run marked. The zoomed-in version (bottom
  panel) clearly demonstrates that at the time of our observations the
nova was at maximum light.}\label{lc}
   \end{figure}

The recent U Sco outburst was announced as an AAVSO\footnote{American Association of Variable Star Observers (AAVSO), Henden, A.A., 2009, Observations from the AAVSO International Database, private communication. http://www.aavso.org/} and VSNet alert on
 2010-Jan-28.438 (UT)\footnote{The 2010 outburst was discovered by two amateur astronomers, Dr. Barbara Harris of New Smyrna Beach and Shawn Dvorak of Clermont, who were part of the AAVSO network} and was confirmed soon thereafter. Our data were obtained
on 2010-Jan-29.354 (UT), about 24 hours after outburst maximum. The AAVSO
light curve (presented in figure~\ref{lc}) indicates that U Sco was at
V$\sim$8 mag the night of our observations. We used the Magellan Echelette (MagE) optical spectrograph on the Clay telescope of the Las Campanas observatories\footnote{http://www.lco.cl/}, with an 1'' slit, providing R$\sim$4,000. The 14 orders of the spectrograph provided effective coverage between 3000 and 9000 $\AA$. During the night of observations, the sky was clear but the moon was high in the sky (albeit setting at the times of our U Sco observations). A total of 50 spectra were acquired sequentially with exposure times of 10 seconds each
and a cadence of minutes. Although the spectrograph is very stable our
object observations were bracketed by observations of a ThAr lamp to
improve our wavelength calibration. Finally, the white dwarf LTT4364
(GJ 430) was observed for flux calibration. For data processing and reduction we used IRAF's\footnote{IRAF is distributed by the National Optical Astronomy Observatories, which are operated by the Association of Universities for Research in Astronomy, Inc., under cooperative agreement with the National Science Foundation} echelle package.

%

   \begin{figure}
   \centering
\includegraphics[angle=0,width=10cm]{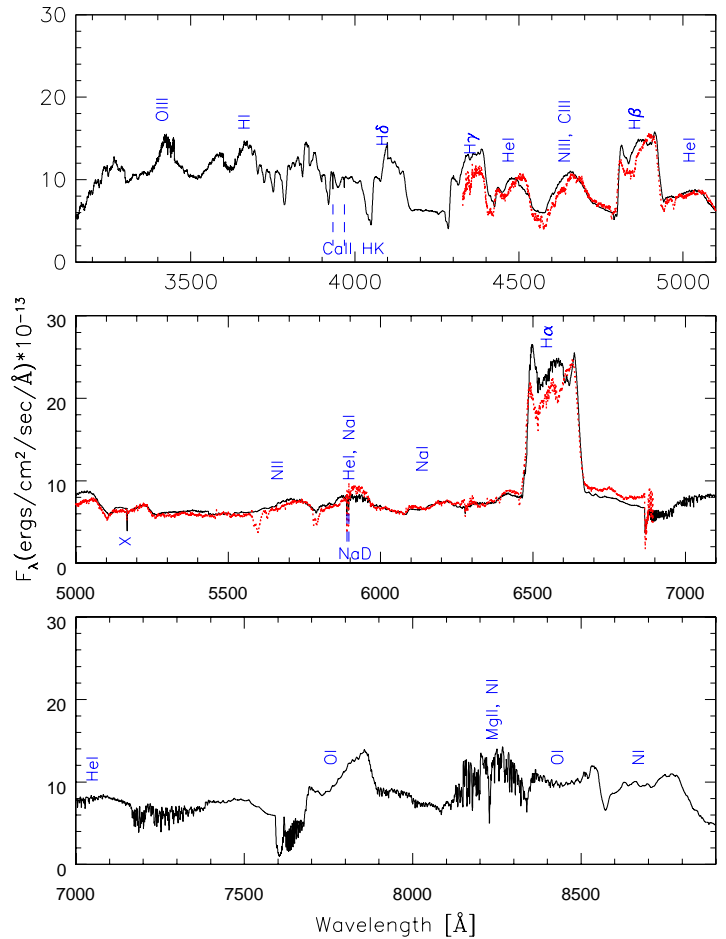}
   \caption{Average MagE spectrum of U Sco, with the main features
  labeled. Overplotted with a red dashed line is the 1999 spectrum from
  \cite{2002AA...387.1013I} for comparison.}\label{spectrum}
   \end{figure}

\section{Data analysis and discussion}

Figure~\ref{spectrum} presents our averaged spectrum of U Sco with the
main features labeled. Using the photometric ephemeris of \cite{1995ApJ...447L..45S},
HJD=2,447,717.6061(32) + 1.2305631(30)E, the observations cover
orbital phases between 0.47 and 0.5, corresponding to inferior
conjunction of the white dwarf. The spectrum is almost certainly formed in the optically thick ejecta of the white
dwarf. Indeed, the spectral features are very strong, full of line blends
and velocity-broadened line profiles. Narrow interstellar Na~I~D and Ca~II~H\&K are the
only absorption lines present that are not associated with the expanding ejecta. The most prominent spectral features are attributed to
N~III, C~III and He~I in the ejecta. They do not show any measurable radial velocity variations within the roughly 2 hr time
period of the MagE observations.

The characteristics of the different outbursts of U Sco seem to be very similar to  each other,  if we
compare the properties  of the various
emission lines, e.g., intensities and line profiles, with those of Anupama and Dewangan (2000) obtained 11 hours
after the 1999 outburst maximum, Iijima (2002) obtained 16 hours after the 1999
outburst maximum, Lepine et al. (1999) obtained 5 days after the 1999
maximum, and Munari et 
al. (1999), obtained 0.64 to 19.8 days after the 1999
 visual maximum. Of those, Iijima (2002) presents spectra only 16
hours after  the outburst  with similar resolution and S/N to the
data presented here, and  Munari et al. (1999)  present multi-epoch
 spectra that explore the time evolution of the Balmer emission lines. Our spectrum is quite similar in its general characteristics to that of Iijima taken the same time after outburst, although there are small differences that are apparent. Figure~\ref{spectrum} includes the \cite{2002AA...387.1013I} spectrum (ref dotted lines) for comparison to ours. The Iijima (2002) spectrum is dominated by a
rich group of N II lines.  The H$\alpha$ line has three main emission
components, with the redder one being the strongest, and multiple satellite absorption features are present at high velocities (reaching -4850km/s). 
The H$\beta$ line is double-peaked, reaching velocities comparable to those of 
 H$\alpha$. The evolution of the 1999 Balmer lines is better
demonstrated in Munari et al (1999), where  the H$\alpha$ line's red peak strengthens
with time and  the line exhibits  three strong peaks 18d after the outburst.

In the 2010 MagE data the line profiles seem to be 
different -- a common phenomenon in postoutburst novae. Both the H$\alpha$ and H$\beta$ lines are triple peaked, and the red and blue peaks reach similar velocities, 
indicating that they originate from the
same part of the ejecta. The H$\gamma$ and H$\delta$ line profiles seem to also have multiple
components which are not resolved due to blending with other
transitions. This is likely the reason for their velocity deviation with respect to the velocities of H$\alpha$ and H$\beta$ in Table~1. The blueshifted absorption components of the H$\alpha$ line identified by \cite{2002AA...387.1013I} are absent in our MagE spectra. A blueshifted emission hump at 6420\AA is also identified by \cite{2002AA...387.1013I} as a secondary component of NII, reaching velocities of
-2880km/s; in the 2010 MagE spectra the same feature is blueshifted by -3056km/s. Multiple high-velocity absorption components from the ejecta are present in the rest of the Balmer lines and their relevant velocities are listed in table~1.

It is interesting to compare the Balmer emission line profiles with those of eclipsing cataclysmic variable accretion disk line profiles, in which the blue and the red anse of the disk produce blueshifted and redshifted components of the Balmer lines with respect to the observer. In the case of U Sco, the red/blue velocities are much higher than those expected from an accretion disk surrounding the WD, so this line shape likely represents the projected geometry of the ejected shell on the orbital plane. The exact shape of the ejected material is not known for U Sco, however modeling of HST and radio observations of the recurrent nova RS Ophiuchi argues for an hourglass shape of the ejecta (\cite{2009ApJ...703.1955R}). It is possible that the U Sco ejecta are also lobe-like, expanding perpendicular to the orbital plane of the binary. Interestingly, the 1999 U Sco outburst spectra of \cite{1999ApJ...522L.121L}, of \cite{1999A&A...347L..39M} and of \cite{2002AA...387.1013I} indicate that over a period of a few days the velocities of those peaks decline. In this case, the central line velocity represents the projected velocity of the ejected blobs as material is accelerated perpendicular to the orbital plane. This would present an obvious similarity of the recurrent nova ejecta with those of bipolar planetary nebulae, suggesting that the explosion dynamics may be  similar. 

\begin{figure}
   \centering
\includegraphics[angle=0,width=8cm]{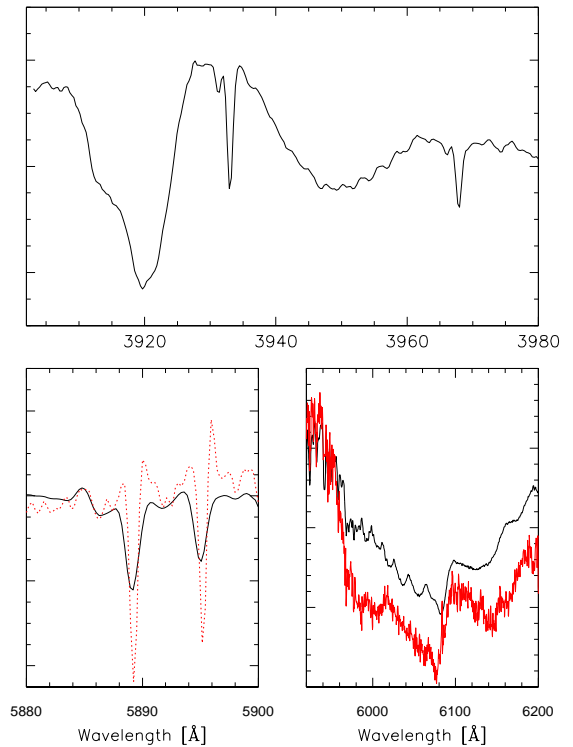}
   \caption{Close-ups of the CaII H\&K line region
  (3920-3980 \AA; top panel) and the region between the HeI 5876~$\AA$
  and Na~I~D and the H$\alpha$ region (5900-6200 \AA; middle
  panel). Overplotted, are the relevant
  spectral regions of the 1999 spectrum from \cite{2002AA...387.1013I}
  for comparison, when available.}\label{lines1}
   \end{figure}

The blue region of the spectrum is dominated by broad emission components of Fe II, O III, He I, C III and various unresolved blends.\cite{2002AA...387.1013I} identifies a suite of N~II lines together with  their various blue/red- shifted components between 5450 and 6400 $\AA$, but  these are not present in our  2010 MagE spectra. Instead, this spectral region is dominated by pronounced wide
emission troughs at 5998, 6012, 6027, 6042, and 6064 \AA (figure~\ref{lines1}, bottom)) likely due to Fe II (at $\lambda\lambda$5997.8, 6012.2, 6027.1, 6042.1, 6065.5 respectively). The NaID lines are the only absorption lines that are present in this wavelength interval. The top panel of figure~\ref{lines1} presents the CaII H\&K lines, demonstrating that each line component is accompanied by two blueshifted absorption dips. The lower velocity components (at -45 and -49 km/s for the Ca~II~H and the Ca~II~K lines respectively) are interstellar in nature, consistent with the low-velocity components of the NaID lines. Note that in 1999 the latter are also accompanied by a red emission component (figure~\ref{lines1}, bottom).

  \begin{figure}
   \centering
\includegraphics[angle=0,width=8cm]{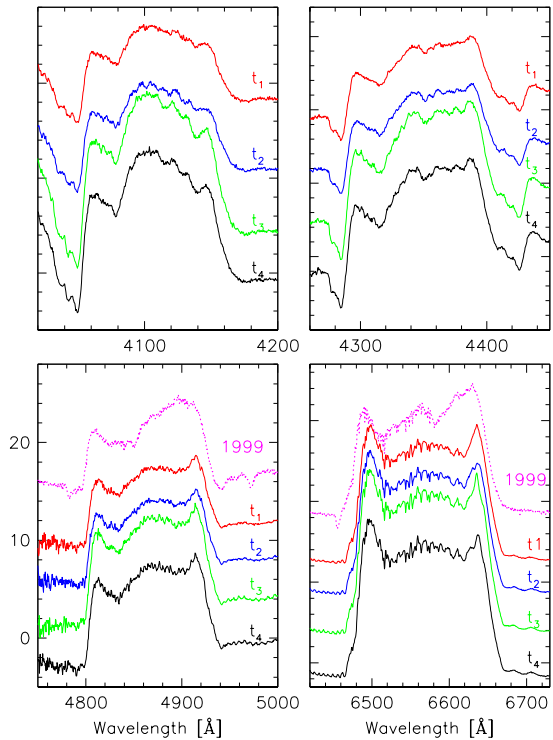}
   \caption{Representative snapshots of Balmer lines: H$\alpha$ (bottom,
  right), H$\beta$ (bottom, left), H$\gamma$ (top, right) and
  H$\delta$ (top, left). t$_2$=t$_1$+672s, t$_3$=t$_1$+1301s and
  t$_4$=t$_1$+1453s. A constant offset was applied to the spectra for
  presentation purposes. Overplotted, are the relevant
  spectral regions of the 1999 spectrum from \cite{2002AA...387.1013I}
  for comparison, when available.}\label{lines2}
   \end{figure}

The medium velocity components \footnote{-186 and -187 km/s for  Ca~II~H and the Ca~II~K lines respectively and -206 and -174 km/sec for NaI D1 and D2 respectively} for both CaII H\&K and NaID are likely to be due to interstellar clouds along the line of sight.
 The high velocity components of the CaII H\&K lines correspond to the velocities of the ejecta. Similar components should be present in th NaI D lines (especially since the NaI ionization potential is much lower at 5.1 eV), however the relevant spectral region is dominated by HeI, FeII, NII and NaI emission which mask possible high velocity absorption components from NaID. 

It is worth noting that variable Na~I~D absorption components from expanding circumbinary gas clouds have been recently detected in the vicinity of three supernovae Ia (SNe~Ia), originating from ionized circumstellar clouds, irradiated by the SN and interacting with the explosion ejecta (\cite{2007Sci...317..924P}). The absence of expected accompanying CaII H\&K lines was attributed to the stronger radiation required to excite CaII H\&K with respect to NaI. In any case, such observations from the SNeIa community are pointing to the single degenerate scenario for SNeIa progenitors. Similar well-documented observations of high-velocity variable NaI and CaII lines in recurrent novae (such as U Sco), present a mechanism for replenishing circumbinary gas with such clouds, providing a bridge between this class of SNe and their progenitors. 

 Finally, at this very early stage in the postoutburst nova there is no indication of forbidden lines or coronal lines in the 2010 U Sco MagE spectra, unlike other recurrent novae, e.g., RS Oph (Rosino 1987), which have giant secondary stars. Furthermore, there are no hints of the weak Mg~I b feature (at 5174$\AA$) and the Fe~II blend (at 5270$\AA$) which were used to secure the spectral classification of the secondary star by \cite{2000AJ....119.1359A}. 

\subsubsection{Line variability}

In a nova explosion the accretion disk surrounding the WD is believed to be swept up by the much more massive ejecta, and the outer layers of the WD are ejected replenishing the surrounding ISM with heavier elements. As discussed before, the velocities of the Balmer emission line components decline with time. In the time-resolved MagE spectra there are no changes in the velocities of the blue and red peaks of the H$\alpha$ and H$\beta$ lines. However, real variability in the line profile {\it shapes} is evident in figure~\ref{lines2}, in which four snapshots of observations of the Balmer lines are presented. Using the notation of the figure, t$_2$=t$_1$+672s, t$_3$=t$_1$+1301s and t$_4$=t$_1$+1453s. More notable variations are presented in the equivalent width of the lines. Although the amplitudes of the variations are not large, they are nevertheless real (within error) and they lack periodicity. The amplitude of those variations is larger in the H$\alpha$ line (of the order of 20$\pm$5$\AA$), declining for  the rest of Balmer lines. The probable cause is variations of the emissivity as the ejecta expand. It is interesting that these fast variations are reminiscent of accretion disk flickering, although the accretion disk should not be visible at the time of the 2010 MagE observations, which is only one day after the peak of the outburst maximum.

\subsubsection{THEA systems?}

Transient Heavy Element Absorption (THEA) systems were described for novae by \cite{2008ApJ...685..451W}, by
examining high-resolution (R$\sim$50,000) optical spectra of
post-outburst nova ejecta. These are transient absorption lines from low ionization Fe-peak and s-process elements that correspond to gas ejected from the secondary star
prior to the nova outburst. The observed gas expansion velocities are between 400-
1000 km/s. The corresponding ions responsible for the absorption are  Sc II, Ti II,
V II, Cr II, Ba II, Sr II, and Fe II (\cite{2008ApJ...685..451W}).   They appear as early as one day after the nova outburst and they evolve with time, having lifetimes
of up to 2 months after the explosion. THEA lines have been identified in 13 novae so
far (\cite{2008ApJ...685..451W}{;} \cite{2010arXiv1004.3600M}), and the exact origin of the gas responsible for the absorption  is still uncertain.

Since the FWHM of THEA lines reaches up to 350~km/s, the 75km/sec
resolution of the MagE U Sco 2010 spectra should be capable of detecting the majority of such systems, if
present, and place constraints on their formation
mechanism. However, they are not observed in the early outburst spectrum of U Sco. 

Thus, U Sco would  join the small group of novae in which THEA are absent (V382 Vel /99, V1187
Sco/04, and V5115 Sgr/05;
\cite{2008ApJ...685..451W}). \cite{2008ApJ...685..451W} point out that
these novae, like U Sco, are all fast novae (t$_{3}$$\le$13d) and therefore the THEA lines may have dissipated
by the time of the relevant observations, which were more than 6-9 days after
outburst. At the same time, if the THEA system is due to material
lingering above the orbital plane of the binary (\cite{2008ApJ...685..451W}) this material could be bipolar and not spherically symmetric, since it is ejected from the secondary star (it mimics an outflow, but it originates from the secondary star of the binary). Therefore, it 
would not be easily detected even in very high resolution spectra in high inclination
systems such as U Sco. Of course, as gas expands from the binary, the opening angle would tend to increase, so components perpendicular to the orbital plane (and further away form the binary) could be detected. Certainly, high resolution observations of U~Sco up to a month after its outburst maximum can provide essential
information on the distribution of the gas responsible for the THEA
absorption lines, and can contribute critically to determining their
formation mechanism. 

 The absence of THEA systems may have implications for outburst models such as those of Hachisu et al. (2000) that predict postoutburst winds, since the winds should produce observable absorption when they achieve a certain column density.  What limits does the absence of observed absorption place on the total mass loss due to such winds?  An answer to this question is provided by the analysis of the THEA system in the nova LMC 2005 by Williams et al. (2008), where for the best case observed until the present time of a narrow line absorption system a Fe II column density of $\sim$10$^{18}$cm$^{-2}$ and a corresponding shell mass of $\sim$10$^{-5}$ M$_{\sun}$ was deduced, assuming solar abundances.  If we assume as an extreme that the LMC 2005 THEA absorption might still have been detectable if it had been 100 times weaker with the same line widths, we could have possibly detected ejected mass of $\sim$10$^{-7}$ M$_{\sun}$, which is roughly what Iijima (2002) estimated for the 1999 U Sco outburst, and is what the Hachisu et al. (2000) models generally predict for early wind mass loss.  However, the velocity gradients in the postoutburst winds are not the narrow 50 km/s THEA widths that were observed for LMC 2005, rather are of order 2,000 km/s for the winds.  Thus, any line absorption from winds is likely to be spread over a much greater wavelength interval, rendering the resulting very broad absorption lines more difficult to detect against the continuum.  For absorption lines formed over a broad velocity interval, roughly 3$\times$10$^{-6}$ M$_{\sun}$ of ejected mass would be needed to detect absorption features from the winds.  Such a mass is greater than the upper limit normally expected from a wind in the early weeks following outburst and therefore the absence of observed THEA absorption should not normally impose a stringent constraint on wind models.  In any event, absorption produced by the outburst ejecta, which are generally more massive, are likely to dominate most absorption from postoutburst winds except in cases of high wind mass loss rates or a situation where the wind continues for many months, because the ejecta and the winds have similar, very large velocity gradients.

The absence of THEA absorption in U Sco within one day of maximum brightness indicates that the circumbinary gas reservoir observed around most novae by Williams et al. (2008) may not be present for this system.  \cite{2010Ap&SS.327..207W} have suggested that nova outbursts may be initiated by two types of mass transfer from the secondary star onto the WD: (1) from steady flow via the inner Lagrangian point, L1, and (2) via a more irregular collapse of the large circumbinary reservoir that produces THEA absorption.  Outbursts of the first type would not necessarily show any THEA absorption and according to TNR models (\cite{2005ApJ...623..398Y}) should be characterized by rapid, smooth declines in visible brightness, especially if these "type 1" outbursts are associated with massive WDs.  We suggest here that the rapid, smooth decline of U Sco is a sign of a TNR triggered by inner Lagrangian mass transfer and the absence of a significant circumbinary reservoir in the line of sight.

\section{Overview of conclusions}

This paper presents the earliest  medium-resolution spectrum of U Sco one day after
its 2010 outburst, covering the optical region of the spectrum
(3000-8500 $\AA$). Wide, multicomponent Balmer emission lines are present, reaching velocities of $\sim$3000km/sec, indicative of an expanding shell resulting from the explosion. Broad emission components of Fe II, O III, He I, C III are identified in the blue part of the spectrum. The THEA complex is not present in U Sco, although the fact that this may be due to our viewing angle of the system can not be ruled out. Time-resolved spectra revealed intrinsic fast variations in the Balmer lines, similar to accretion disk flickering.

\begin{table*}
\caption{Kinematic characteristics of the main line profiles in the U Sco spectrum }
\begin{tabular}{lcccccccc}
\hline
Line & rest wavelength & \multicolumn{3}{c}{Velocities of line
components$^{\mathrm{a}}$} & EW$^{\mathrm{b}}$ & FWHM$^{\mathrm{c}}$ & em/abs & notes \\
& (\AA) & Blue & Red & center of line & (\AA) &  (\AA)& & \\
&   & (km/sec)     & (km/sec)   & (km/sec) & & & &  \\
\hline
H$\alpha$&  6563  &  -2915  &  3283  &  150  &  -309$\pm$5  &  109 & em &ejecta \\
NaD1& 5896 &-44 &--  &--  &  0.231$\pm$0.003  &  1.11 & abs & IS\\
&&-206 &--  &--  &  0.044$\pm$0.002  &  1.34 & abs & IS\\
NaD2& 5890 &-45&--   &--  & 0.313$\pm$0.003  &  1.15 & abs & IS\\
&&-174 & --  &--  &0.016$\pm$0.003  &  0.85 & abs & IS\\
H$\beta$&  4874  &  -2966  &  3287  &  -390  &  -145$\pm$5  &  129 & em &ejecta \\
& &-3854 &--  &--   &--  &--  & abs & ejecta \\
& &-5864 &--  & -- & -- &--  &  abs & ejecta\\
H$\gamma$ &  4342  &  --  &  --  &  584  &  -138$\pm$6  &  99 & em &ejecta \\
& &-3806 & -- &  -- & -- &  --& abs & ejecta \\
& & -5864 & -- &  -- & -- & -- & abs & ejecta \\
H$\delta$ & 4100  &  --  &  --  &  370 &  -110$\pm$10   &  87 &em &ejecta \\
& &-3786  & -- & -- & -- & -- & abs & ejecta \\
& &-4304  & -- & -- & -- & -- & abs & ejecta \\
CaII~H & 3968 &-45  &-- &--  & 0.115$\pm$0.012  &  0.937& abs & IS\\
& & -187&--  &--  & 0.009$\pm$0.002  &  0.618 & abs &IS\\
& &-1547 & -- &--   & 2.340$\pm$0.050  &  14.92 &abs & ejecta\\
CaII~K & 3934 & -49 & --  &--  &0.181$\pm$0.023  &  0.902 & abs & IS\\
& &-187 &--  &--  &  0.018$\pm$0.005  &  0.617 & abs & IS\\
& & -1106$^{\mathrm{d}}$&--  &--   & 3.580$\pm$0.060  &  10.32 & abs & ejecta\\
\hline
\end{tabular}
\begin{list}{}{}
\item[$^{\mathrm{a}}$]The blue/red velocities in the line profiles
  presented here were measured from the average spectrum with a
  Gaussian fit; the average error in the velocities is 2km/sec. The
  velocities of the central components of the Balmer lines
  weremeasured ith IRAF/splot's task ``e'', providing an average error
  of 5km/s.
\item[$^{\mathrm{b}}$]Measured from the average spectrum. Following the conventional IRAF nomeclature, negative EW values correspond to emission lines. The errors in the measurements correspond to uncertainties in the determination of the line continuum.
\item[$^{\mathrm{c}}$]Measured from the average spectrum; the error in FWHM measurements is less than 1$\AA$ in all cases.
\item[$^{\mathrm{d}}$]Blend with H$\epsilon$ components.
\end{list}
\end{table*}

\begin{acknowledgements}
We would like to thank Kevin Krisciunas for notifying us about the new U Sco eruption and T. Iijima for making his 1999 U Sco spectrum
available to us. We would also like to thank our anonymous referee for the careful reviewing of the manuscript. SK acknowledges support by NASA's NAI Program. 
We acknowledge with thanks the variable star observations from the AAVSO International Database contributed by observers worldwide and used in this research.
\end{acknowledgements}

\end{document}